\documentclass{article}

\usepackage{hyphenat}
\usepackage{chngpage}
\usepackage{amsmath}
\usepackage{xspace}
\usepackage{mathrsfs}
\usepackage{graphicx}
\usepackage{algorithmic}
\usepackage{algorithm}
\usepackage{url}
\usepackage{algorithm, algorithmic}
\floatname{algorithm}{Algorithm}
\usepackage{booktabs}
\usepackage[usenames,dvipsnames]{color}
\usepackage{lipsum}
\usepackage{afterpage}
\usepackage{array}

\usepackage{enumitem}
\setlist{nosep}

\newcommand{\MD}{\operatorname{M2}}
\newcommand{\MQ}{\operatorname{M4}}
\renewcommand{\AA}{\operatorname{AA}}
\newcommand{\EE}{\mathcal{E}}
\newcommand{\algoname}{LlamaFur\xspace}

\usepackage{amssymb}
\usepackage{authblk}

\begin{document}
\bibliographystyle{plain}

\date{}
\title{\algoname: Learning Latent Category Matrix \\ to Find Unexpected Relations
in Wikipedia}

\author[1]{Paolo Boldi\thanks{Partially supported by the EU-FET grant NADINE (GA 288956).}}
\author[1]{Corrado Monti}
\affil[1]{Dipartimento di Informatica\\
Universit\`a degli Studi di Milano\\
Italy\\
\texttt{\{paolo.boldi,corrado.monti\}@unimi.it}
}

\maketitle
\begin{abstract}
Besides finding trends and unveiling typical patterns, modern information
retrieval is increasingly more interested in the discovery of surprising
information in textual datasets. In this work we focus on finding
\emph{unexpected links} in hyperlinked document corpora when documents are
assigned to categories. To achieve this goal, we model the hyperlinks graph
through node categories: the presence of an arc is fostered or discouraged by
the categories of the head and the tail of the arc. Specifically, we determine a
latent category matrix that explains common links. The matrix is built using a
margin-based online learning algorithm (Passive-Aggressive~\cite{CDKSSOPAA}),
which makes us able to process graphs with $10^{8}$ links in less than $10$
minutes.
We show that our method provides better accuracy than most existing text-based
techniques, with higher efficiency and relying on a much smaller amount of
information. It also provides higher precision than standard link prediction,
especially at low recall levels; the two methods are in fact shown to be
orthogonal to each other and can therefore be fruitfully combined.
\end{abstract}

%

\section{Introduction}

In general, data mining (text mining, if the data involved take the form of
textual documents) aims at extracting potentially useful information from
some (typically unstructured, or poorly structured) dataset.
The basic and foremost aim of data mining is discovering frequent patterns, and
this problem attracted and still attracts a large part of the research efforts
in this field.
Nonetheless a quite important and somehow dual problem is that of finding
unexpected (surprising, unusual, new, unforeseen\dots) information; it is
striking that this line of investigation did not receive the same amount of
attention. 

Albeit there is some research on the determination of surprising information in
textual corpora (most often based on the determination of outliers in the
distribution of terms or $n$-grams) there is essentially no work dealing with
\emph{unexpected links}. Even if some of the previous proposals exploiting text
features can be adapted to this case, a simpler (and, as we here show, more
effective) way to approach this problem is by using link prediction
algorithms~\cite{LZLPCNS}, stipulating that a link that is difficult to predict
is unexpected.

In this paper, we prove that the availability of some form of categorization of
documents can significantly improve the techniques described, leading to
algorithms that are ex\-treme\-ly efficient, use much less information than
text-based methods, and offer better precision/recall trade-offs. Compared to
link prediction, our technique also provides higher precision at low recall
levels; moreover, the two methods have orthogonal outputs, and therefore their
combination improves over both.

Our idea is that if the documents within a linked corpora are tagged with
categorical information, one can learn how category/category pairs influence the
presence of links, and as a consequence determine which links are unusual (in
the sense that they are not ``typical''). For example, documents of the category
``Actor'' often contain links to documents of the category ``Movie'' (simply
because almost all actor pages contain links to the movies they acted in). The
fact that George Clooney used to own Max, a 300-pound pig, for 18 years presents
itself as a link from an ``Actor'' page to a page belonging to the category
``Pigs''/``Coprophagous animals'', which is atypical in the sense above.

Our basic algorithm -- henceforth called \emph{\algoname}, ``Learning LAtent
MAtrix to Find Unexpected Relations'' -- tries to learn a category/category
matrix describing the latent relations between categories: to this aim we adapt
a Passive-Aggressive learning algorithm, since it scales very well with the size of
data, making us able to process $10^{8}$ links in less than $10$ minutes. Then
we reconstruct those links that are explainable according to the
matrix, and those that cannot be justified by the categories alone. Not only
\algoname is also more efficient than both link prediction and the previous
techniques based on the analysis of the textual content of the page, but it also
improves the accuracy of link prediction algorithms in identifying unexpected
links, if the two are combined.

\smallskip
It is worth noting that the discovery of unexpected links offers a chance to
find unknown information: given a certain document, we can highlight text
snippets containing unexpected links. Meaningful text is often characterized, in
web documents, by the presence of links that enrich its semantic; this is
especially true in the case of Wikipedia, often used as a knowledge base for
ontologies. Its link structure has proven to be a powerful resource for many
tasks~\cite{SKWYAGO,PSDLSTW}. For this reason, finding unexpected links seems a
valuable way to detect meaningful text with information unknown to the reader.

We wish to remark that our results could be in principle applied to a plethora
of different kinds of objects. The only assumption on the input is a (possibly
directed) graph, and a meaningful categorization of its nodes; categories can be
overlapping as well, so in fact they may just be some observed features of each
object.
These assumptions are quite general, and could be useful in many real-world use
cases, from the detection of unexpected collaborations between grouped
individuals (e.g., consider a co-authorship network, with authors categorized by
their past and present affiliations) to finding surprising travel habits from
geo-tagged data. Other possible applications include finding unusual patterns
for fraud detection~\cite{coderre1999fraud} and data forensics~\cite{forensics}.

The paper is organized as follows. In Section~\ref{sec:related-work} we will
review other works dealing with mining of unexpected information. Our technique
will be presented in Section~\ref{sec:learningcatmat}, where we explain how
to estimate a latent category matrix through online learning;
Section~\ref{sec:naive}, where we describe a more naive way to compute it,
which will be used as a baseline; in Section~\ref{sec:using-catmat} we
show how to use the category matrix to measure the unexpectedness of a link. In
Section~\ref{sec:experiments} we exhibit experimental evidence for the efficacy of our methods, by comparing them
with different approaches derived from literature. Finally, in
Section~\ref{sec:conclusions} we will sum up our work and suggest possible
directions for future research.

\section{Related Work}
\label{sec:related-work}

One of the first papers trying to consider the problem in the context of text
mining was~\cite{LMYDUICWS}. In that work, two supposedly similar web sites are
compared (ideally, two web sites of two competitors). The authors first try to
find a match between the pages of the two web sites, and then propose a measure
of unexpectedness of a term when comparing two otherwise similar pages. All
measures are based on term (or document) frequencies; unexpected links are also
dealt with but in a quite simplistic manner (a link in one of the two web sites
is considered ``unexpected'' if it is not contained in the other). 
Note that finding unexpected information is crucial in a number of contexts,
because of the fundamental role played by serendipity in data mining (see,
e.g.,~\cite{Ramakrishnan1999,Murakami2007}).

This unexpectedness measure is taken up in~\cite{JLDUDC}, where the aim is that
of finding documents that are similar to a given set of samples and ordering the
results based on their unexpectedness, using also the document structure to
enhance the measures defined in~\cite{LMYDUICWS}. Finding outliers in web
collections is also considered in~\cite{ABAHAWCOMWQV}, where again dissimilarity
scores are computed based on word and $n$-gram frequency.  

Some authors approach the strictly related problem of determining lacking
content (called \emph{content hole} in~\cite{NAATMCHSCTC}) rather than
unexpected information, using Wikipedia as knowledge base. A similar task
is undertaken by~\cite{EAYHWCHWSICWW}, this time assuming the dual approach of
finding content holes in Wikipedia using the web as a source of information.

More recently,~\cite{TOYITDUIBPUACOTR} considers the problem of finding
unexpected related terms using Wikipedia as source, and taking into account
at the same time the relation between terms and their centrality.

An alternative way to approach the problem of finding unexpected links is by
using \emph{link prediction}~\cite{LZLPCNS}: the expectedness of a link $e$ in
a network $G$ is the likelihood of the creation of $e$ in $G-\{e\}$. In fact, we
will later show that state-of-the-art link prediction algorithms
like~\cite{AAFNW} are very good at evaluating the (un)expectedness of links.
Nonetheless, it turns out that the signal obtained from the latent category
matrix is even better and partly orthogonal to the one that comes from the graph
alone, and combining the two techniques greatly improves the accuracy of both.

\smallskip
Our basic idea -- explaining a graph according to some feature its nodes
exhibit -- is not new. It has been proposed as a graph model by some authors,
and it is often called \emph{latent feature model}. Features can be real-valued --
as in~\cite{latent-factor-models} -- or binary, as in our case, where the set of
nodes exhibiting a feature is sharp, and not fuzzy. These models have been
studied by different authors (such as~\cite{MGJ2009}), either with known or
unknown features. 

Such models have been successfully applied to social networks. They were studied
as \emph{affiliation networks}~\cite{lattanzi_affiliation} by Lattanzi and
Sivakumar; in that work, a social graph is produced by a latent bipartite
network of \emph{actors} and \emph{societies}. Kim and Leskovec have presented a
model called \emph{multiplicative attribute graph}~\cite{leskovec-mag}, where
the network is a result of each node attributes. Their model is based on a
different $2 \times 2$ matrix for each attribute, and it can describe complex
behaviour between categories, such as homophily and heterophily. They show
empirical evidence of its ability to explain real-world networks.

We will present a model based on a single latent category-category
matrix, for the particular scope of mining unexpected links in the graph.
Indeed, we will show how our model is able to capture the notion of
\emph{surprising} and \emph{unsurprising} relations. Furthermore, thanks to its
simplicity, it is easy and fast to infer the latent category network from data.

The same problem -- explaining links through features of each node -- can be
casted as a Latent Dirichlet Allocation (LDA)~\cite{blei2003latent} problem; usually,
in this context, features are the words contained in each document. For
example,~\cite{liu2009topic} and~\cite{chang2009relational} build a link
prediction model obtained from LDA, that considers both links and features of
each node. However, the largest graphs considered in these works have about
$10^3$ nodes (with $\sim10^4$ possible features), and they do not provide running
time. \cite{henderson2009applying}~developed an LDA approach explicitly tailored
for ``large graphs'' --- but without any external feature information for nodes;
the largest graph they considered has about $10^4$ nodes and $10^5$ links, for
which they report a running time of $45-60$ minutes. The algorithm we propose,
although simpler, requires 9 minutes to run on a graph three orders of magnitude
larger (about $10^6$ nodes and $10^8$ links).

Interpreting links in a network as a result of features of each node has a solid
empirical background. The simple phenomenon of homophily -- i.e, links between
nodes sharing the same features -- has been widely studied in social
networks~\cite{mcpherson2001birds} and other complex
systems~\cite{homophilymedia}. More complex behavior, where nodes with certain
features tend to connect to another type of nodes, has also proven to be greatly
beneficial in analyzing real social networks. Tendencies of such kind are called
\emph{mixing patterns} and are often described by a category-category matrix~\cite{crimaldi}.
For example, they appeared to be a crucial factor in tracking the spread of
sexual diseases~\cite{aral1999sexual} as well as in modelling the transmission
of respiratory infections~\cite{mossong2008social}. For this reason, such
matrices are also called ``Who Acquires Infection From Whom'' (WAIFW) matrices,
and have been empirically assessed in the field through
surveys~\cite{hens2009mining} and with wearable sensors~\cite{isella2011close}.


\section{Learning the category matrix}
\label{sec:learningcatmat}

Consider a directed graph $G=(D,L)$ (the ``document graph''), whose nodes $d\in
D$ represent \emph{documents} and whose arcs $(d,d') \in L$ represent
\emph{(hypertextual) links} between documents. Further assume that we have a set
$C$ of \emph{categories} and that each document $d\in D$ is assigned a set of
categories $C_d \subseteq C$. 

Our first goal is to reconstruct the most plausible latent ``category matrix''
that explains the observed document graph; more precisely, we wish to find a
$C \times C$ real-valued matrix $W$ such that
\begin{equation}
\label{eqn:sums}
	\sum_{c \in C_{d}} \sum_{c' \in C_{d'}} w_{c,c'}
\end{equation}
is positive iff $(d,d') \in L$. This relation implicitly defines a graph model:
given each category set $C_{d}$ and the latent category matrix $W$, one can obtain a graph.

We are going to assume that in most cases a relation is \emph{unexpected} --
that is, surprising to the reader -- if it is poorly explained by a plausible
category matrix. We will put this assumption under test in the experimental
section.

To find such a matrix $W$, we recast our goal in the framework of online binary
classification. As we will explain later on, the idea here is that by learning
how to separate links from non-links, the classifier must infer $W$ as its
internal state. Binary classification, in fact, is a well-known problem in supervised
machine learning. Suppose to have a training set of \emph{examples}, each
example $\mathbf{x}_{i}$ associated with a binary label
$\widehat{y_{i}}\in\lbrace-1,1\rbrace$; based on these data, the problem is to
build a classifier able to label correctly unknown data. \emph{Online}
classification simplifies this problem by assuming each example is presented in
a sequential fashion; the classifier (1) observes an example; (2) tries to
predict its label; (3) receives the true label, and consequentially updates its
internal state; (4) moves on to the next example. An online learning algorithm,
generally, needs a constant amount of memory with respect to the number of
examples, which allows to employ these algorithms in a situation where a very
large set of voluminous input data is available -- like in our case. 

A well-known type of online learning algorithms are the so-called
perceptron-like algorithms. They all share these traits: each example must be a
vector $\mathbf{x}_{i}\in\mathbb{R}^{n}$; the internal state of the classifier
is also represented by a vector $\mathbf{w}\in\mathbb{R}^{n}$; the predicted
label is $y_i=sign(\mathbf{w} \cdot \mathbf{x}_{i} )$, and the algorithms differ
on how $\mathbf{w}$ is built. Perceptron-like algorithms (for example, ALMA and
Passive-Aggressive) are usually simple to implement, provide tight theoretical
bounds, and have been proved to be fast and accurate in
practice~\cite{GenALMA,CDKSSOPAA}. For these reasons, we will reduce our problem
to online binary classification.

To this aim, let us represent each document $d$ with the indicator vector of
$C_d$, i.e., with the binary vector $\mathbf{d}$ such that $d_c=1$ \emph{iff} $c
\in C_{d}$. Now, an \emph{example} will be a pair of documents $(d, d')$,
represented as the outer product kernel $\mathbf{d} \otimes \mathbf{d'}$: this
is a matrix where the element $[\mathbf{d} \otimes \mathbf{d'}]_{c,c'}$ is $1$
iff the first document belongs to $c$ and the second to $c'$. This $(|C| \times
|C|)$-matrix\footnote{In practice, we normalize this matrix so that it has unit
$L1$-norm. Normalization often gives better results in
practice~\cite{cristianini2000introduction}; in this case, documents belonging
to few categories provide stronger signals than those that belong to many
categories.} can be alternatively thought of as a vector of size $|C|^2$,
allowing us to use them as training examples for a perceptron-like classifier,
where the label is $y=1$ iff $(d,d') \in L$ (if there is a link), and $y=-1$
otherwise. The learned vector $\mathbf{w}$ will be, if seen as a $|C| \times
|C|$ matrix, the desired $W$ appearing in (\ref{eqn:sums}). In other words, we
are using $|C|^2$ features, in fact a kernel projection of a space of dimension
$2|C|$ onto the larger space of size $|C|^2$. Similarly the weight vector to be
learned has size $|C|^2$. Positive examples are those that correspond to
existing links.

\paragraph{A Passive-Aggressive algorithm}

Among the existing per\-cep\-tron-like online classification frameworks, we
chose the well-known Passive-Aggressive classifier, characterized by being
extremely fast, simple to implement, and shown by many
experiments~\cite{CARVALHO2006SINGLE, MRZZAC} to perform well on similar
datasets. To cast this algorithm for our case, let us consider a sequence of
pairs of documents
\[ 
	(d_1,d_1'),\dots,(d_T,d_T') \in D^2
\]	
(to be defined later).
Define a sequence of matrices $W_0,\dots,W_T$ and of slack variables $\xi_1,\dots,\xi_T \geq 0$
as follows:
\begin{itemize}
  \item $W_0=0$
  \item $W_{t+1}$ is a matrix minimizing $\|W_{t+1}-W_t\|+K\xi_{t+1}$ subject to
  the constraint that 
  \begin{equation}
  \label{eqn:ineq}
  \sigma(d_t,d_t') \cdot \sum_{c \in C_{d_t}} \sum_{c' \in C_{d_t'}}
  w_{t+1}(c,c')\geq 1 - \xi_{t+1},  
  \end{equation}
  where 
  \[
  	\sigma(x,y)=\begin{cases}
  					-1	& \text{if $(x,y) \not\in L$}\\
  					1	& \text{if $(x,y) \in L$}
  				\end{cases},
  \]
  $\|-\|$ denotes the Frobenius norm and $K$ is an optimization parameter
  determining the amount of aggressiveness.
\end{itemize}
The intuition behind the above-described optimization problem~\cite{CDKSSOPAA}
is the following:
\begin{itemize}
  \item the left-hand-side of the inequality (\ref{eqn:ineq}) is positive iff
  $W_{t+1}$ correctly predicts the presence/absence of the link $(d_t,d_t')$;
  its absolute value can be thought of as the confidence of the prediction;
  \item we would like the confidence to be at least 1, but allow for some error
  (embodied in the slack variable $\xi_{t+1}$);
  \item the cost function of the optimization problem tries to keep as much
  memory of the previous optimization steps as possible (minimizing the
  difference with the previous iterate), and at the same time to minimize the
  error contained in the slack variable.
\end{itemize} 

By merging the Passive-Aggressive solution to this problem with our
aforementioned framework, we obtain the algorithm described in
Alg.~\ref{alg:our-pa-alg}.

\begin{algorithm}[t]
		\textsc{Input}: \\
		\-\hspace{0.4cm} The graph $(D, L)$, with $L \subseteq D \times D$ \\
		\-\hspace{0.4cm} Categories $C_d \subseteq C $ for each document $d \in D$
		\\
		\-\hspace{0.4cm} A parameter $K>0$ \\
		\textsc{Output}: \\
		\-\hspace{0.4cm} The latent category matrix $W$ \\
		\begin{enumerate}
			\item $W \leftarrow \mathbf{0}$
            \item Let $(d_1,d_1'),\dots,(d_T,d_T')$ be a sequence of
                  elements of $D \times D$.
			\item For $i=1,\dots,T$
			
			\begin{enumerate}
				\item $\rho \leftarrow \frac{1}{ |C_{d_i}| \cdot |C_{d_i'}| }$
				\item $\mu \leftarrow \sum_{c \in C_{d_i} }{ \sum_{c' \in C_{d_i'}} {
				w_{c,c'} }}$
				\item \textbf{If} $(d_i,d_i') \in L$ \\
						\-\ $\quad \delta \leftarrow \rho \cdot \min(K, 1-\mu \rho) $ \\
					  \textbf{else} \\
						\-\ $\quad \delta \leftarrow - \rho \cdot \min(K, 1+\mu \rho) $ \\
				\item For each $c \in C_{d_i}$,  $c' \in C_{d_i'}$: \\
						\-\ $\quad w_{c,c'} \leftarrow w_{c,c'} + \delta $
			\end{enumerate}
		\end{enumerate}
\caption{\label{alg:our-pa-alg}Passive-Aggressive algorithm to build the latent category matrix.}
\end{algorithm}


Please note that our aim is not to build a perfect classifier: instead, we will
use this algorithm to find a plausible category-category matrix. The model found
by the classifier will be used later to detect outliers (as described for
example in~\cite{AGGARWAL2013OUTLIER}).

\paragraph{Sequence of pairs} In our case, $W$ is built through a
single-pass online learning process, where we have all positive examples at our
disposal (and they are in fact all included in the training sequence), but where
negative examples cannot be all included, because they are too many and they
would produce overfitting. 

The Passive-Aggressive construction described above depends crucially on the
sequence of positive and negative examples $(d_1,d_1'),\dots,(d_T,d_T')$ that is
taken as input. In particular, as discussed
in~\cite{JAPKOWICZ2002CLASSIMBALANCE}, it is critical that the number of
negative and positive examples in the sequence is balanced. Taking this
suggestion into account, we build the sequence as follows: nodes are enumerated
(in arbitrary order), and for each node $d \in D$, all arcs of the form
$(d,-)\in E$ are put in the sequence, followed by an equal number of pairs of
the form $(d,-)\not\in E$ (for those pairs, the destination nodes are chosen
uniformly at random). Of course, if $m=|E|$ is the number of links, then $T=2m$
and the sequence contains all the $m$ links along with $m$ non-links.

Obviously, there are other possible ways to define the sequence of examples and
to select the subset of negative examples. We suggest some of them in
Section~\ref{sec:conclusions}. However, we chose to adopt this technique --
single pass on a balanced random sub-sample of pairs -- in order to test our
methodology with a single, natural and computationally efficient
approach.\footnote{We carried out experiments performing more passes on the same
subsample; it slightly increased (less than $2\%$) the \emph{accuracy} of $W$ --
i.e., the number of pairs that are correctly classified. However, it is dubious
whether the increased time cost is worth the limited improvement in terms of
unexpectedness mining.}

\section{A naive way to build the category matrix}
\label{sec:naive}
  Let us describe an alternative, naive variant of how the latent
  category matrix $W$ could be obtained. Recall that the purpose is to use
  equation (\ref{eqn:sums}) to compute the expectedness of a link $(d,d')$.
  With this purpose, we shall use a naive Bayes
  technique~\cite{BishopPatternRec}, estimating the probability of existence of
  a link through maximum likelihood and assuming independence between
  category memberships.
  
  For a given category $c$, let $D_c$ be the set of documents that have the
  category $c$; let also $\EE_{c,d}$ represent the event that $d$ belongs to the
  category $c$ (i.e., $c \in C_d$ or, equivalently, $d \in D_c$). Now for any
  two categories $c$ and $c'$ one can compute the probability that there is a
  link between two documents that belong to those categories as
  \[
  	p_{c,c'}=P[(d,d') \in L \mid \EE_{c,d} \text{ and } \EE_{c',d'}].
  \]
  This quantity can be naively estimated as the fraction of pairs $(d,d')$ such
  that $\EE_{c,d} \wedge \EE_{c',d'}$ that happen to be links. In other
  words,
  \[
  	p_{c,c'}=\frac{|\{(D_c \times D_{c'}) \cap L\}|}{|D_c|\cdot |D_{c'}|}.
  \]
  For a specific pair of documents $(d,d')$, the probability of the presence of 
  a link is given by
  \[
  	P\left[(d,d')\in L\biggl.\biggr| \bigcap_{c \in C_d} \EE_{c,d} \text{ and
  	}\bigcap_{c' \in C_{d'}} \EE_{c',d'}\right].
  \]
  Now, under some independence assumptions\footnote{More precisely, we are
  assuming that $\EE_{c,d}$ and $\EE_{c',d'}$ are independent, whenever $c \neq
  c'$ or $d \neq d'$, and also that they are independent even under the
  knowledge that $(d,d')\in L$.}, the latter can be expressed as
  \begin{multline*}
  	\prod_{c \in C_d} \prod_{c' \in C_{d'}} P\left[(d,d')\in L\biggl.\biggr| \EE_{c,d} \text{ and }
  	\EE_{c',d'}\right]= \\
  	=\prod_{c \in C_d} \prod_{c' \in C_{d'}}  p_{c,c'}.
  \end{multline*}
  Applying a logarithm and add-one smoothing~\cite{russellnorvig},
  this is rank-equivalent to
  \[
  	\sum_{c \in C_d} \sum_{c' \in C_{d'}} w_{c,c'}
  \]
  where
  \[
  	w_{c,c'}=\log \frac{|\{(D_c \times D_{c'}) \cap L\}| + 1 }{(|D_c|+1)\cdot (|D_{c'}| + 1)}
  \]
  This is yet another way to define the matrix $W$ used in the \algoname
  algorithm; the resulting expectedness score for link $(d,d')$ is given by
  (\ref{eqn:sums}), and will be referred to as Naive-\algoname.

\section{Using the category matrix}
\label{sec:using-catmat}

\begin{figure}[ht]
	\begin{center}
		\begin{tabular}{cc}
			\includegraphics[scale=0.15]{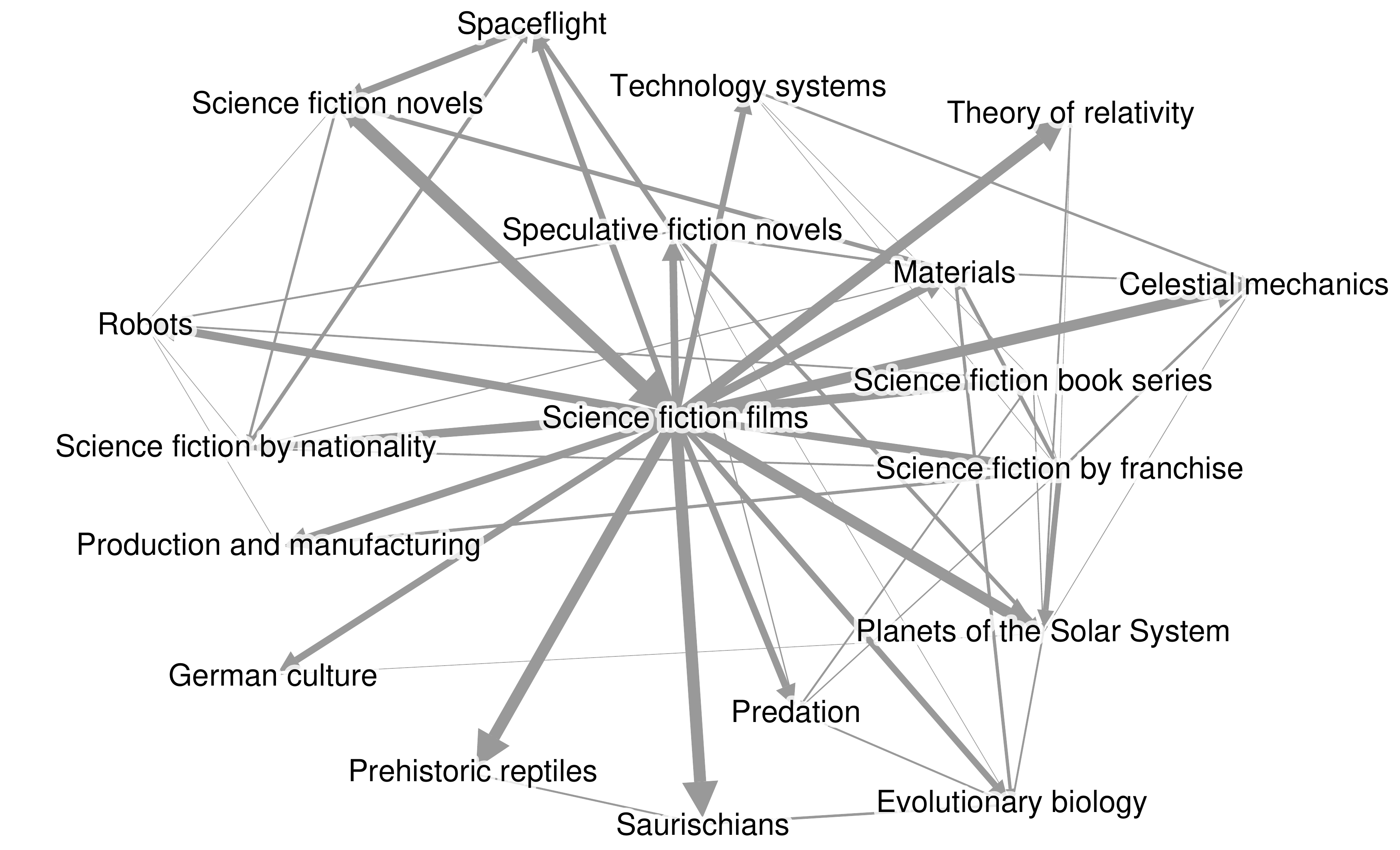}
			&
			\includegraphics[scale=0.15]{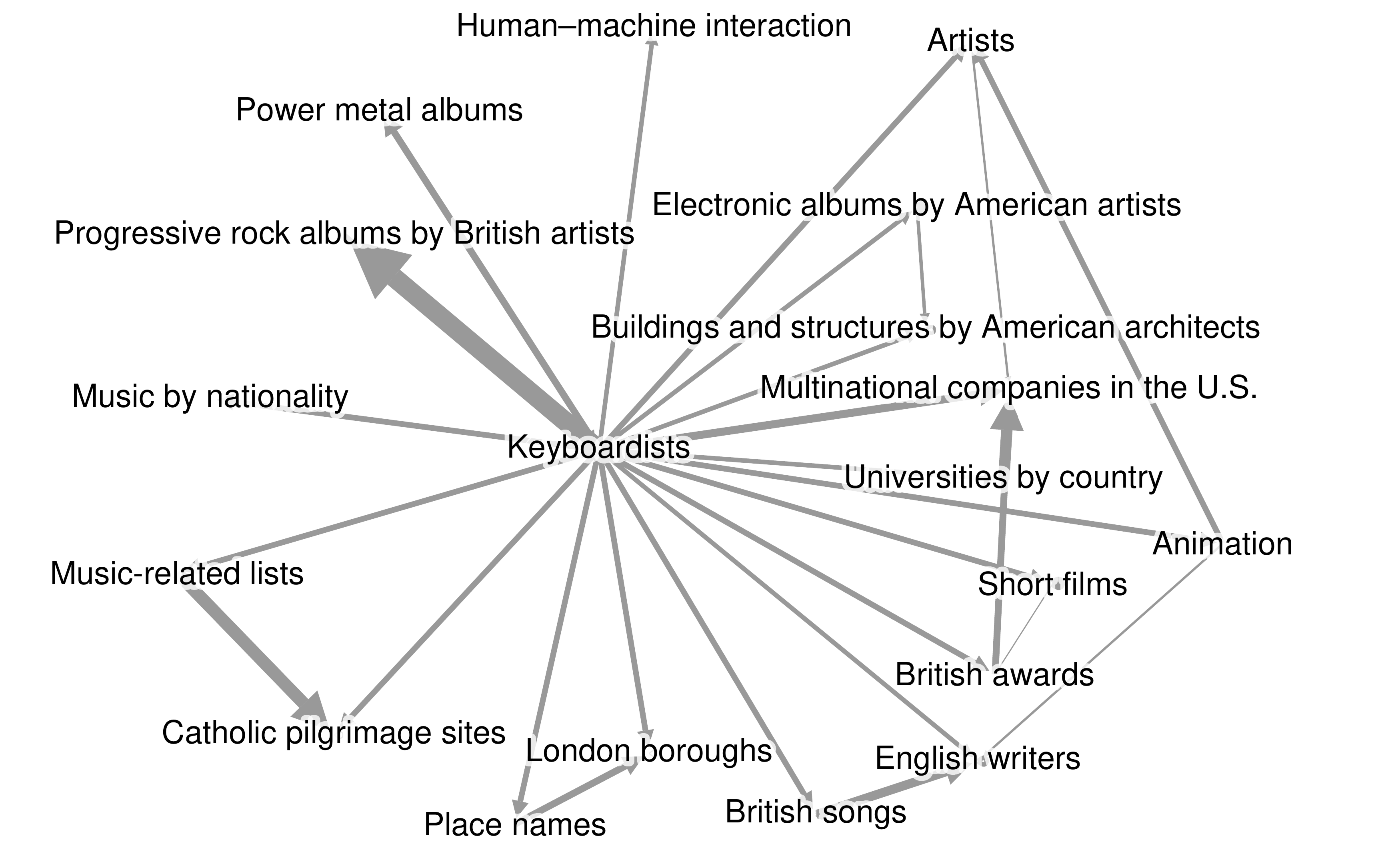}
		\end{tabular}
		\caption{\label{fig:category-networks}
            Two fragments of the latent category graph induced by \algoname
            matrix $W$, representing the 18 closest neighbors of categories
            ``Science Fiction Films'' and ``Keyboardists'', respectively.
			The width of the arc from $c$ to $c'$ is proportional to
			$w_{c,c'}$, and arcs with $w_{c,c'}\leq 1$ are not shown.}
    \end{center}
\end{figure}

Let us now call $W$ the category matrix obtained at the end of the learning
process (that is, $W=W_T$, according to the notation of
Section~\ref{sec:learningcatmat}), or equivalently the matrix built using the
naive approach of Section~\ref{sec:naive}. This matrix allows one to sort the
links $(d,d')\in L$ in increasing order of $\sum_{c \in C_{d_t}} \sum_{c' \in
C_{d_t'}} w_{c,c'}$ (i.e., by increasing explainability): the first links are
the most unexpected.

In particular, in the case of the learning approach of
Section~\ref{sec:learningcatmat}, one can build a graph $G^*=(D,L^*)$ whose
links are the set $L^*$ of pairs $(d,d')$ such that
\[
	\sum_{c \in C_{d_t}} \sum_{c' \in C_{d_t'}} w_{c,c'} > 0.
\]

In a standard binary-classification scenario, $G^*$ would be the graph $G$ that
our classifier learned. In particular, the elements of the set $L\setminus L^*$
($L^* \setminus L$, resp.) are the false negative (false positive, resp.)
instances. 

But ours is \emph{not} a link-prediction task, and we do not expect in any sense 
that $L$ and $L^*$ are similar. In particular, we shall
certainly observe a phenomenon that we can call \emph{generalization effect}:
suppose that it frequently happens that a document assigned to a category $c$
(e.g., an actor) contains links to documents assigned to another category $c'$
(e.g., a movie). This will probably make $w_{c,c'}$ very large, and so we may
falsely deduce that \emph{every} document assigned to $c$ (every actor) contains
a link to \emph{every} document assigned to $c'$ (every movie).  In other words,
\algoname\ cannot be used as a reliable link-prediction algorithm. 

The generalization effect will, by itself, make $L^*$ much larger than $L$
(i.e., it will produce many false positive instances), but we do not care much
about this aspect. Our focus is \emph{not} on trying to reconstruct $L$, but
rather in understanding which elements of $L$ are difficult to explain based on
the categories of the involved documents. We say that a link $(d,d')\in L$ is
\emph{explainable} iff $(d,d')\in L^*$; the set of explainable links is
therefore $L\cap L^*$. On the contrary, the elements of $L \setminus L^*$ are
called unexplainable, and these are the links we want to focus on. 

\begin{figure}[ht]
	\begin{center}
		\begin{tabular}{cc}
			\includegraphics[scale=0.15]{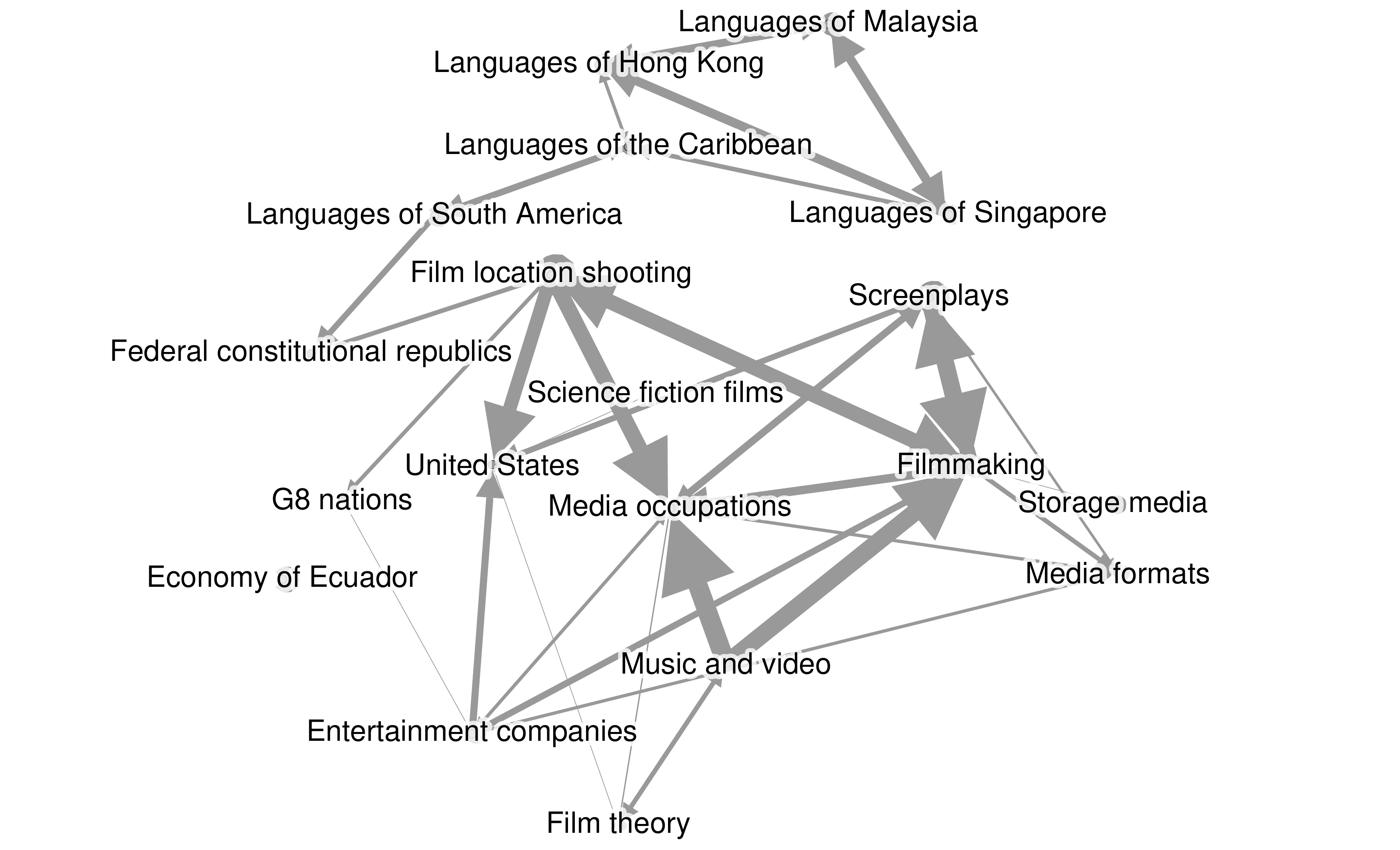}
			&
			\includegraphics[scale=0.15]{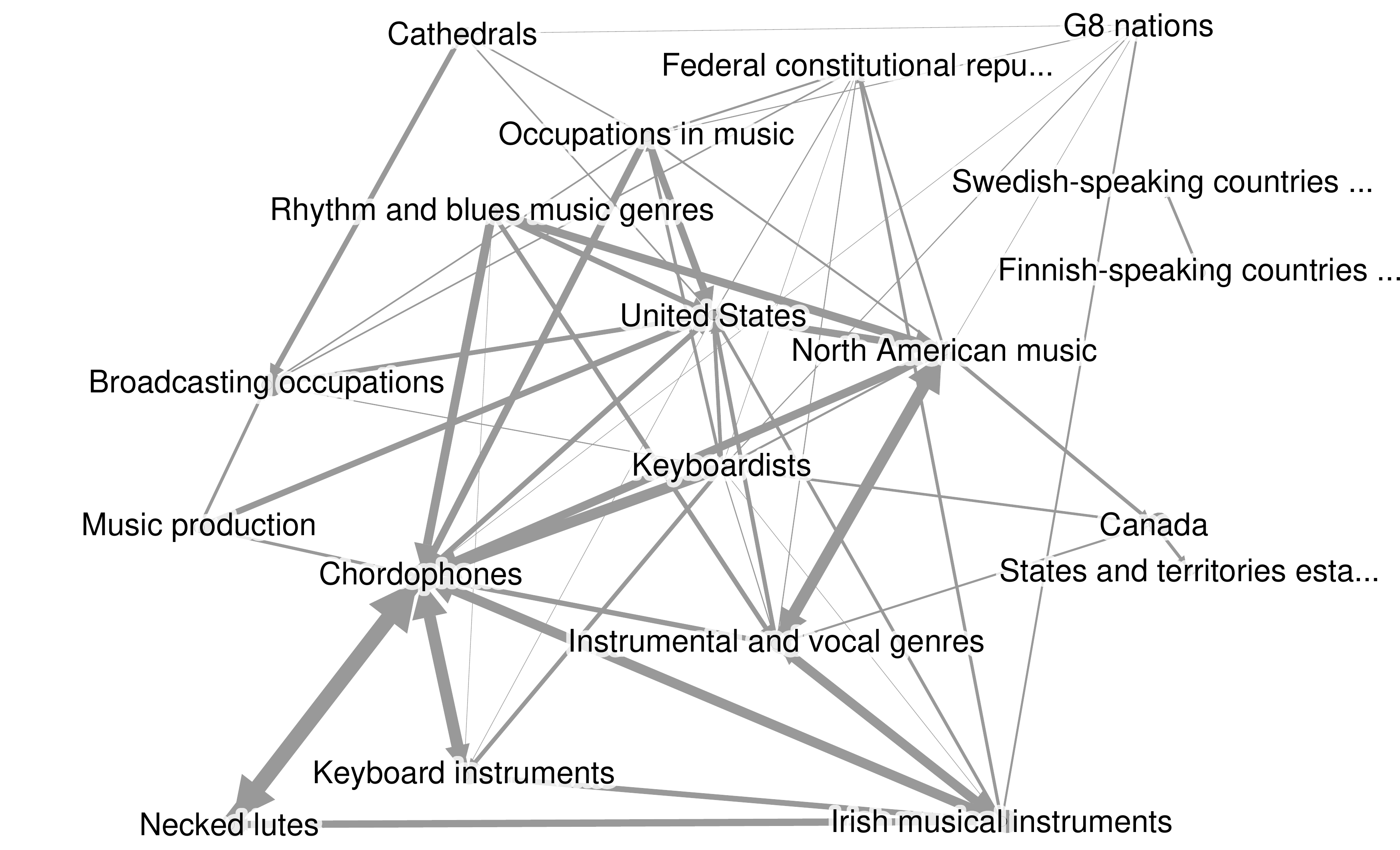}
		\end{tabular}
		\caption{\label{fig:naive-category-networks}
            Two fragments of the latent category graph induced by Naive-\algoname
            matrix, representing the 18 closest neighbors of categories
            ``Science Fiction Films'' and ``Keyboardists'', respectively.
			The width of the arc from $c$ to $c'$ is proportional to
			$w_{c,c'}$, and the lighter arcs are not shown. For comparison
			with \algoname, see Figure~\ref{fig:category-networks}.}
    \end{center}
\end{figure}

In Figure~\ref{fig:category-networks} we show two small examples of how the
matrix $W$ learned as in Section~\ref{sec:learningcatmat} looks like, when
considering the Wikipedia dataset (for a full explanation of how the dataset
was built, see Section~\ref{sec:experiments}): in the picture, we display the
18 neighbours closer to two starting categories (``Science Fiction Films'' and
``Keyboardists''); the width of the arc from $c$ to $c'$ is proportional to
$w_{c,c'}$, and arcs with $w_{c,c'}\leq 1$ are not shown. For example, from the
picture it is clear that a link from a page of a science-fiction film to a page
of a science-fiction novel is highly expected, as it is one from a page of a
keyboardist to one of a British progressive rock album. 
From this representations we can catch a glimpse of how this method is able to
build a model for the graph, capturing meaningful relations
between categories.
The rougher version of the same neighborhood as induced by Naive-\algoname is
shown in Figure~\ref{fig:naive-category-networks}: even from this small 
example, it is clear that the naive version introduces more noise (epitomized by
the inclusion of ``Language of the Carribean'' and ``Languages of Singapore''
among the 18 closest neighbors of ``Science fiction films'').

\section{Experiments}
\label{sec:experiments}

Given its increasing importance in knowledge
representation~\cite{SKWYAGO}, we used the English edition of Wikipedia as
our testbed. In particular, we employed the \texttt{enwiki}
snapshot\footnote{This dataset is commonly referred to as
\texttt{enwiki-20140203-pages-articles} according to Wikipedia naming scheme.}
of February 3, 2014 to obtain:

\begin{itemize}
	\item the document graph, composed by $4\,514\,662$ Wikipedia pages, with as many as
	$110\,699\,703$ arcs; every redirect was merged to its target page; 
	\item the full categorization of pages: a map associating every page to one
	of the $1\,134\,715$ categories;
	\item the category pseudo-tree: a graph built by Wikipedia editors, with the aim of assigning each category to a ``parent'' category.
\end{itemize}

\paragraph{Wikipedia categories}
The first problem is that the categorization on Wikipedia is quite noisy and, in
fact, a continuous work-in-progress: categories may contain only one
(or even no) page, they might be duplicates of each other, and so on. 
We therefore cleansed the page categorization using the methods described
in~\cite{BoMCWCC}; in a nutshell, we computed the
harmonic centrality measure~\cite{BoVAC} on the category hierarchy, and
considered only the set $C$ of the $20\,000$ most central categories (called
\emph{milestones}).
We chose $20\,000$ as a good compromise between the significance of these categories
(in terms of the F-Measure of Alg.~\ref{alg:our-pa-alg}, see below) and the
space needed to store $W$ (see Fig.~\ref{fig:number-of-categories}).

\begin{figure}[htb]
		\includegraphics[width=1.0\columnwidth]{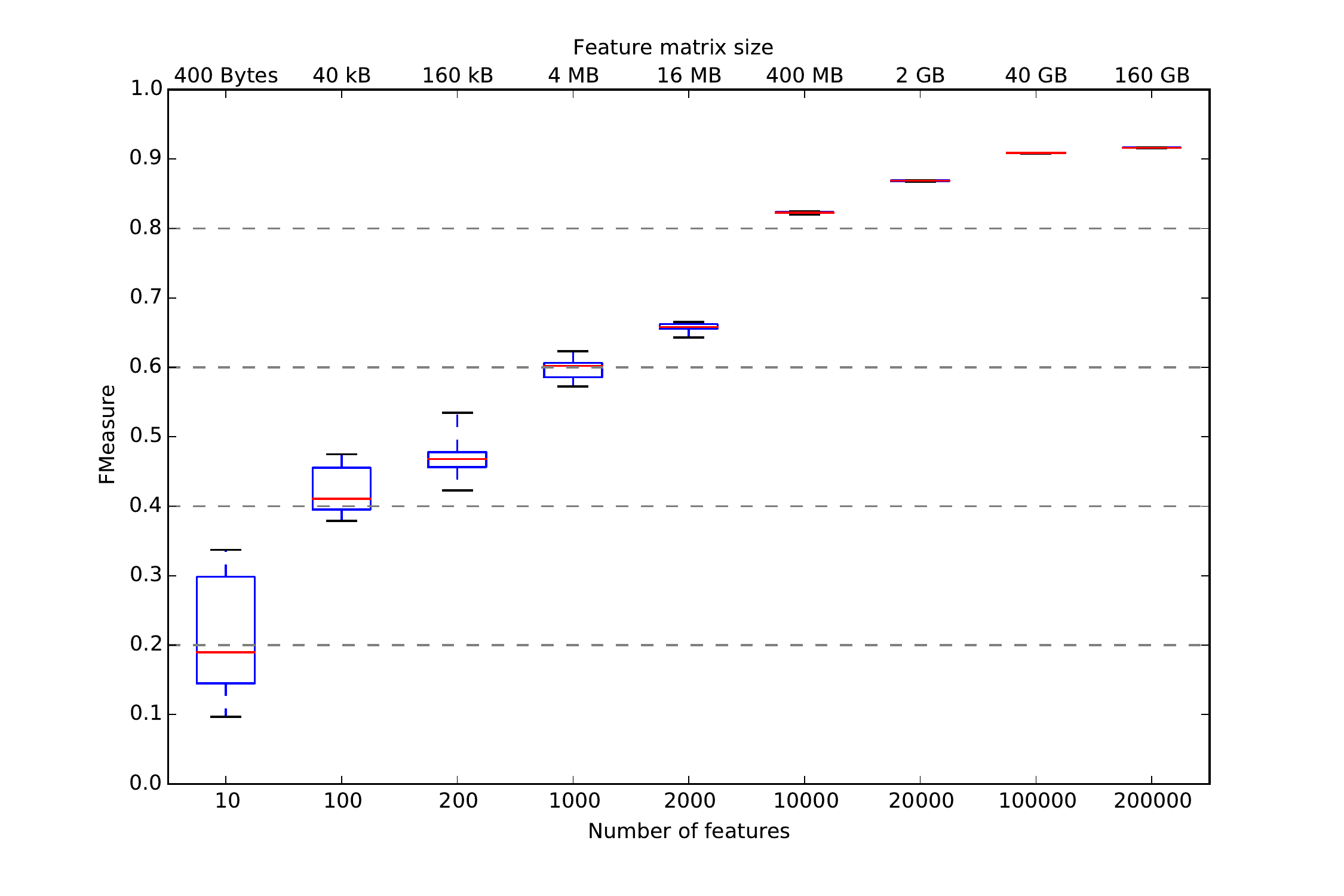}
		\caption{\label{fig:number-of-categories}
			F-Measure of Alg.~\ref{alg:our-pa-alg} in 10-fold cross-validation, with different number of considered categories. For each cardinality, we report the size required to store a dense representation of $W$.
		}
\end{figure}

We then computed, for every category $c$, the milestone $\iota(c)\in C$
closest to $c$ in the pseudo-tree (distances are computed as shortest paths in the
pseudo-tree), and re-categorized all the pages applying $\iota(-)$ to its
original categories. If there is no $c' \in C$ connected to $c$, $\iota(c)$ is undefined and we simply discarded $c$. 

At the end of the day, we obtained a set $C$ of $20\,000$ milestone categories,
and a map associating each Wikipedia page $d$ to $C_d \subseteq C$; on average, each page belongs to
$4$ categories. This cleansing process yields
very clean labels; for instance, pages originally belonging to the category
``Swiss manuscripts'' are now categorized in ``Swiss culture'', and similarly 
``Elections in Southwark'' is remapped to ``Local government in London'', 
``Flamenco compositions'' to ``Spanish music'', etc.
Since the method we present is based on a
well-defined categorization of labels, reducing the noise in the labels may be a key factor
behind our results.

\smallskip
We proceeded then to apply \algoname to extract the latent category
matrix $W$. In doing so, we first used a 10-fold cross-validation technique to
assess how well Alg.~\ref{alg:our-pa-alg} is generalizing; specifically, we
divided in 10 folds the space of node pairs $D \times D$. The results,
reported in Table~\ref{tab:crossvalidation}, are consistent with our
expectations, and suggest that we can consider the unexplained links as
atypical. In particlar, the average of F-Measure on unknown node pairs is
$86.9\%$, showing that the model learnt by \algoname is robust and not
threatened by overfitting.

\begin{table}
	\begin{center}
    \normalsize
		\begin{tabular}{l|c}
            \textbf{Measure} & \textbf{Results} \\ \hline
			Accuracy 	& $87.3\% \pm 0.06$ \\ 
			Precision 	& $90.0\% \pm 0.05$ \\
			Recall 	    & $84.0\% \pm 0.14$ \\
			F-Measure 	& $86.9\% \pm 0.07$ \\ 
		\end{tabular}
		\caption{\label{tab:crossvalidation}
                Average and standard deviation of results from a 10-fold
                cross-validation of Alg.~\ref{alg:our-pa-alg} on the links of
                Wikipedia.
			}
    \normalsize
	\end{center}
\end{table}

Learning on the whole graph, the ratio $|L\cap L^*|/|L|$ -- that is,
how many existing links are explained by $W$ -- is equal to $86\%$. Please note
that running our algorithm on the whole Wikipedia graph ($110\,699\,703$ arcs)
and the categorization we chose ($20\,000$ categories) required only 9 minutes
on an Intel Xeon CPU with 2.40GHz. We illustrated previously in
Fig.~\ref{fig:category-networks} some fragments of $W$.~Finally, we proceeded to
assign our unexpectedness score to each link. An example of the largest and lowest
scores for two Wikipedia articles is provided in Table~\ref{table:anectodical}
and~\ref{table:anectodical2}.

\begin{table}[htb]
    \scriptsize
    \begin{tabular}{p{2cm}p{3.5cm}|p{2cm}p{3.5cm}}
        \multicolumn{4}{c}{\large{Links of \textbf{Jupiter}}} \\ \hline
        \multicolumn{2}{c|}{\textit{\normalsize{\textbf{Most Unexpected}}}} &
        \multicolumn{2}{c}{\textit{\normalsize{\textbf{Most Expected}}}} \\
        \hline
            \textbf{Inquisition} &    
			\emph{[Observation of Jupiter moons]} was a major point in
                favor of Copernicus' heliocentric theory of the motions of the
                planets; Galileo's outspoken support of the Copernican theory
                placed him under the threat of the Inquisition.
            &
            \textbf{Galileo Galilei} &
                Galilean moons were first discovered by Galileo Galilei in
                1610.
        \\            
        \hline
			\textbf{Proto-Indo-European language} &    
                \emph{[Jupiter]} name comes from the Proto-Indo-European
                vocative compound *Dyeus-pater (nominative: *Dyeus-pater,
                meaning "O Father Sky-God", or "O Father Day-God").
			&
			\textbf{E. E. Barnard} &
                In 1892, E. E. Barnard \emph{[an American astronomer]} observed
                a fifth satellite of Jupiter with the 36-inch (910 mm) refractor
                at Lick Observatory in California.
         \\
		\hline
            \textbf{ Gan De } &    
                Gan De, a Chinese astronomer, made the discovery of one of
                Jupiter's moons in 362 BC with the unaided eye.
			&            
			\textbf{Ptolemy} &
                \emph{[Ptolemy]} constructed a geocentric planetary model based
                on deferents and epicycles to explain Jupiter's motion relative
                to the Earth.
           \\
         \hline
            \textbf{ Fish } &    
                In 1976, before the Voyager missions, it was hypothesized that
                ammonia or water-based life could evolve in Jupiter's upper
                atmosphere. This hypothesis is based on the ecology of
                terrestrial seas which have simple photosynthetic plankton at
                the top level, fish at lower levels feeding on these creatures,
                and marine predators which hunt the fish.
			&            
            \textbf{Jupiter (mythology)} &
                The Romans named the planet after the Roman god Jupiter.
    \end{tabular}
    \caption{\label{table:anectodical} Most expected and most unexpected links
    in the Wikipedia article \emph{Jupiter}, according to \algoname.
	}
    \normalsize
\end{table}

\begin{table}[htb]
    \centering
    \scriptsize
    \begin{tabular}{p{2cm}p{3.5cm}|p{2cm}p{3.5cm}}
        \multicolumn{4}{c}{\large{Links of \textbf{Kim Jong-il}
            \footnotesize{\emph{(supreme leader of North Korea from 1994 to 2011)}} }} \\ \hline
        \multicolumn{2}{c|}{\textit{\normalsize{\textbf{Most Unexpected}}}} &
        \multicolumn{2}{c}{\textit{\normalsize{\textbf{Most Expected}}}} \\
        \hline
            \textbf{Elvis Presley} &    
                In a 2011 news story, The Sun reported Kim Jong-il was obsessed
                with Elvis Presley. His mansion was crammed with his idol's
                records and his collection of 20,000 Hollywood movies included
                Presley's titles -- along with Rambo and Godzilla. He even copied
                the King's Vegas-era look of giant shades, jumpsuits and
                bouffant hairstyle.
			&            
            \textbf{George W. Bush} &
                Kim's regime argued the secret \emph{[nuclear]} production was
                necessary for security purposes -- citing the presence of United
                States-owned nuclear weapons in South Korea and the new
                tensions with the United States under President George W. Bush.
            \\
	        \hline
			\textbf{Michael Jordan} &    
                Kim reportedly enjoyed basketball. Former United States
                Secretary of State Madeleine Albright ended her summit with Kim
                by presenting him with a basketball signed by NBA legend Michael
                Jordan.
            &
            \textbf{Kim Il-sung} &
                He succeeded his father and founder of the DPRK, Kim Il-sung.
			\\
	        \hline
            \textbf{Sonbong} &    
                \emph{[Kim Jong-il's]} father returned to Pyongyang that
                September, and in late November Kim returned to Korea via a
                Soviet ship, landing at Sonbong.
            &
            \textbf{Adolf Hitler} &
                A report \emph{[...]} concluded that the ``big six" group of
                personality disorders shared by dictators Adolf Hitler, Joseph
                Stalin, and Saddam Hussein were also shared by Kim Jong-il.
			\\        
        	\hline
			\textbf{Korean Air Flight 858} &    
                South Korea accused Kim of ordering the 1983 bombing in Rangoon,
                Burma which killed 17 visiting South Korean officials, including
                four cabinet members, and another in 1987 which killed all 115
                on board Korean Air Flight 858.
            &
            \textbf{Malta} &
                Kim is also said to have received English language education at
                the University of Malta in the early 1970s.
    \end{tabular}
    \caption{\label{table:anectodical2} Most expected and most unexpected links
    in the Wikipedia article \emph{Kim Jong-il}, according to \algoname.
	}
    \normalsize
\end{table}

\paragraph{Evaluation methodology}
We want to evaluate the effectiveness of \algoname in mining unexpected links
using a standard approach commonly adopted in Information Retrieval. In our
context, a \emph{query} is a document, the possible \emph{results} are the
hyperlinks that the document contains, and a result is \emph{relevant} for our
problem if it represents an unexpected link. The scenario we have in mind is
that of a user wishing to find surprising links in a certain Wikipedia page.
What we are trying to assess is how well LlamaFur can identify an unknown set of
unexpected links, having full knowledge of graph and categorization of nodes.

In order to compare the results obtained by
\algoname with the existing state-of-the-art for similar problems, we performed
a user study based on the same pooling method adopted for many standard
collections such as TREC (\url{trec.nist.gov}): we considered a random sample of
237 queries (i.e., Wikipedia documents); for each query we took, among its $t$
possible results (i.e., links), the top-$\left \lfloor{\alpha \cdot t}\right
\rfloor$ most unexpected ones according to each system under comparison (see
below); all the resulting links were evaluated by human beings. We set $\alpha =
0.1$, and obtained about $3\,698$ links. 

The human evaluators were asked to categorize each link into one of four classes
(``totally expected'', ``slightly expected'', ``slightly unexpected'' and
``totally unexpected''). They were provided with the first paragraph of the two
Wikipedia pages, and a link to the whole article if needed. The
resulting dataset of $3\,698$ evaluated links is available for
download\footnote{\url{http://git.io/vmChm}} and inspection. 
(To obtain a sufficiently large number of evaluated links in a short amount of
time, there was virtually no overlap between the links that the evaluators
worked on; however, we manually inspected the dataset to find that
the labels produced are quite robust---we invite the reader to do the same.)
After the human
evaluation, we only considered the queries that have at least one irrelevant
(``totally/slightly expected'') and one relevant (``totally/slightly
unexpected'') result according to the evaluation, obtaining a dataset with $117$
queries. In this dataset, on average each query has $3.48$ relevant results over
$20.9$ evaluated links. About $58.1\%$ of the links were labelled as
``totally expected", only $2.2\%$ were ``totally unexpected" and about
$8.8\%$ were labelled as ``unexpected".

It is evident how unexpected links are very sparse (many pages did not present any
unexpected link at all); this motivated us to employ bpref~\cite{bpref}, a
well-suited information retrieval measure. To compute it, we followed TREC
specifications\footnote{\url{http://trec.nist.gov/pubs/trec16/appendices/measures.pdf}};
in a nutshell, bpref computes an index of how many judged relevant documents
are retrieved ahead of judged non-relevant documents.

\paragraph{Baselines and competitors}
In our comparison, \algoname is tested in combination and against a number of
baselines and competitors. In particular, we considered \algoname and its
naive variant, Naive-\algoname, along with some of the other (un)expectedness
measures proposed in the literature.

Albeit there are, at the best of our knowledge, no algorithms specifically
devoted to determining unexpected links, we can adapt some techniques used for
unexpected documents to our case. All of those methods try to measure the
unexpectedness of a document $d$ among a set of retrieved documents $R$. In our
application, we are considering a link $(d',d)$ and taking $R$ to be the set of
all documents towards which $d'$ has a hyperlink.

\begin{itemize}
  \item \emph{Text-based methods}. In the literature, all of the measures of
  unexpectedness are based on the textual content of the document under
  consideration.
  	\begin{itemize}
  	\item The first index, called
  $\MD$ in~\cite{JLDUDC} (a better variant of $M1$, the measure proposed
  in~\cite{LMYDUICWS}), is defined as:
  \[
  	\MD(d)=\frac{\sum_{t}U(d,t,R)}{m}
  \]
  where $m$ is the number of terms in the dictionary, and $U(d,t,R)$ is the
  maximum between 0 and the difference between the normalized term frequency of
  term $t$ in document $d$ and the normalized term frequency of $t$ in
  $R$ (the set of all retrieved documents). The normalized term frequency is
  the frequency of a term divided by the frequency of the most frequent term.
  \item The second index, called $\MQ$ in~\cite{JLDUDC} (where they prove that
  it works better than $\MD$ in their context), is the
  \[
  	\MQ(d)=\max_{t} \operatorname{tf}(t) \cdot \log
  	\frac{|R|}{\operatorname{df}(t)}
  \]
  where $\operatorname{tf}(t)$ is the normalized term frequency of term $t$ in
  $d$, and $\operatorname{df}(t)$ the number of documents in $R$ where $t$
  appears.
  	\end{itemize}
  \item \emph{Link-prediction methods}. A completely different, alternative
  approach to the problem is based on \emph{link prediction}: how likely is
  it that the link $(d',d)$ is created, if we assume that it is not there? 
  Among the many techniques for link prediction~\cite{LZLPCNS}, we 
  tested the well-known \emph{Adamic-Adar index}~\cite{AAFNW} ($\AA$, in
  the following), defined\footnote{The formula is applied to the symmetric
  version of the graph, in our case; note that this (like \algoname) is a
  measure of expectedness, whereas $\MD$ and $\MQ$ are measures of
  unexpectedness.} by
  \[
  	\AA(d,d')=\sum_{d'' \in \Gamma(d) \cap \Gamma(d')}
  	\frac1{\log |\Gamma(d'')|},
  \]
  where $\Gamma(d)$ is the set of documents which $d$ links to.
  \item \emph{Combinations}. Besides testing all the described techniques in
  isolation, we tried to combine them linearly. Since each unexpectedness
  measure exhibits a different scale, we first need to normalize each measure by
  taking its \emph{studentized residual}\footnote{The
  (internally) studentized residual is obtained by dividing the
  residual (i.e., the difference from the sample mean)
  by the sample standard deviation.}~\cite{CWRIR}. It is worth mentioning here
  that~\cite{latent-factor-models,MGJ2009} can also be used as link-prediction
  algorithms, and would apparently fit better our approach because they predict
  links based on binary node features, but the crucial difference is that in those
  models features are \emph{latent} and need to be reconstructed, whereas in our
  scenario features (categories) are readily available.
\end{itemize}

\paragraph{Results}

In the following, we are only going to discuss the best algorithms and
combinations, besides some of the most interesting alternatives. The raw average
bpref values are displayed in Table~\ref{tab:avgbpref}.
%
%

\begin{table}[htb]
	\begin{center}
		\begin{tabular}{l|c|c}
			{\bf Algorithm} & {\bf Average bpref} & {\bf Input data}\\
			\hline
			$\AA$     			& 0.286 & graph				\\
			$\MD$   			& 0.179 & bag of words 		\\
			$\MQ$   			& 0.293 & bag of words 		\\
			Naive-\algoname 	& 0.251 & graph, categories \\
			\algoname 			& 0.343 & graph, categories	\\
			\algoname + $\AA$ 	& 0.350 & graph, categories	\\
		\end{tabular}
		\caption{\label{tab:avgbpref}
				Average values for bpref.
			}
	\end{center}
\end{table}

\begin{figure}[htb]
		\includegraphics[width=1.1\columnwidth]{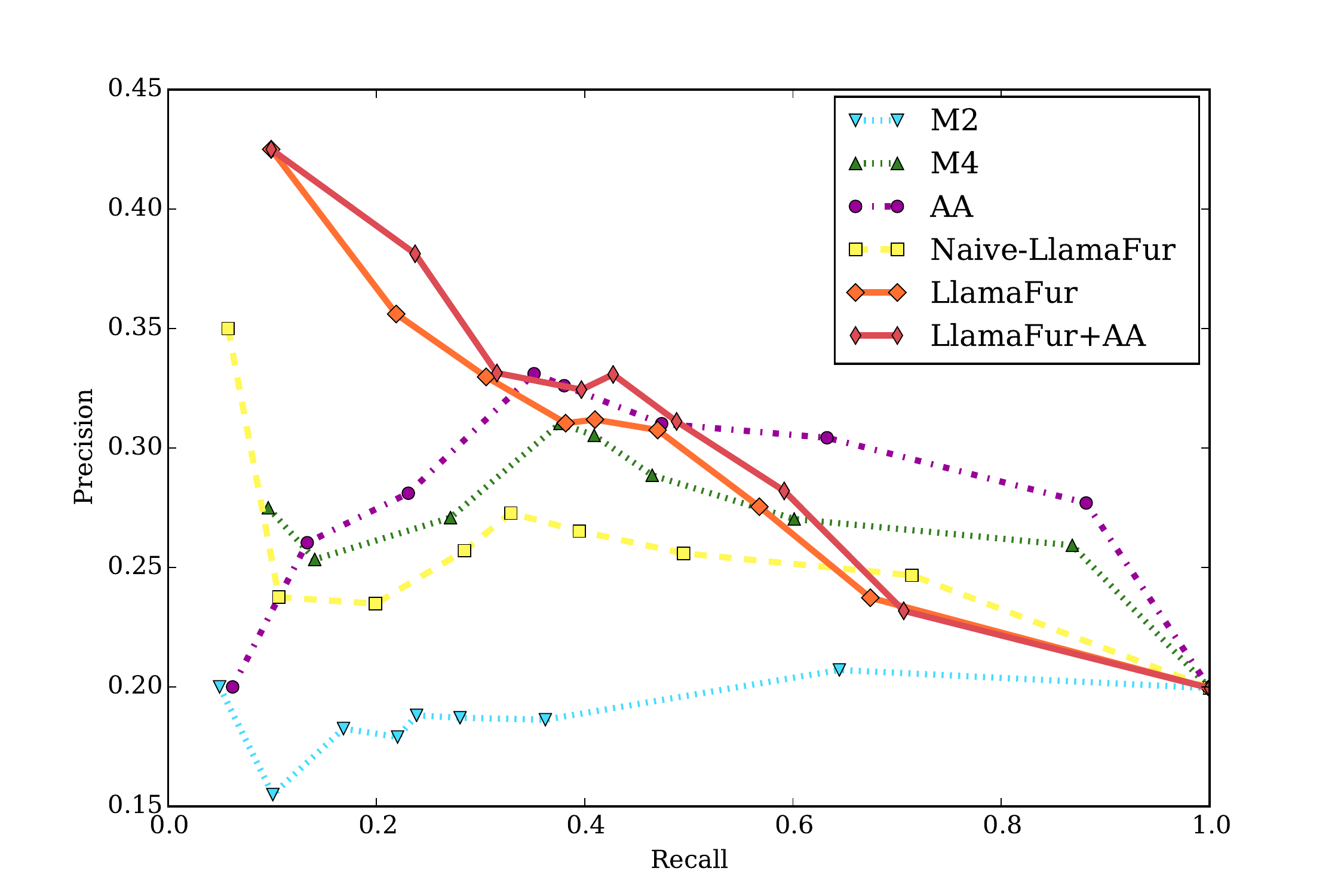}
		\caption{\label{fig:precision-recall}
			Average precision-recall values evaluated after the
			$1$st, $2$nd, $5$th, $8$th, $10$th, $15$th, $25$th, $50$th, and $100$th percentiles
			for each query.
		}
\end{figure}

\begin{figure}[htb]
	\begin{center}
		\includegraphics[width=1.1\columnwidth]{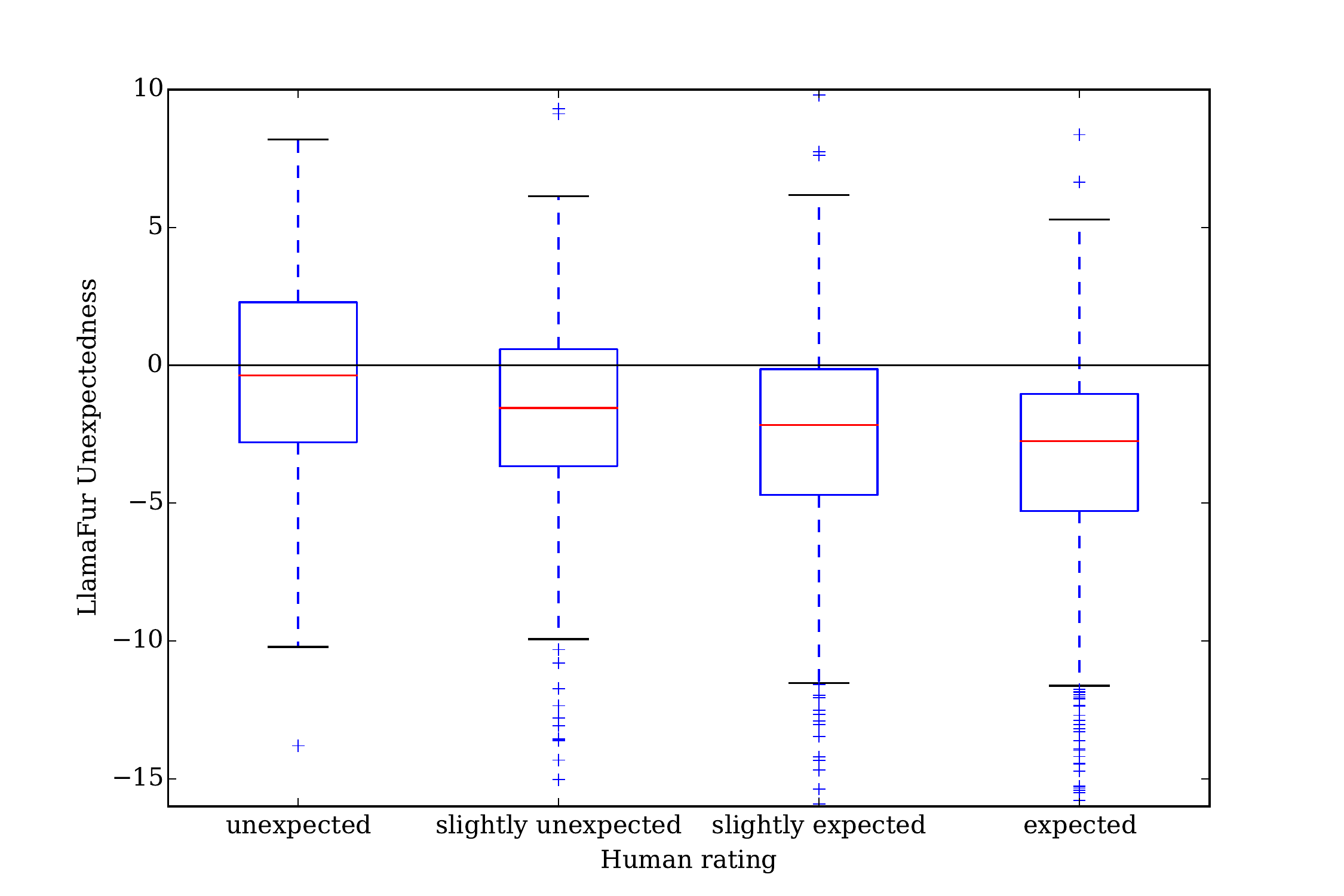}
		\caption{\label{fig:rating-vs-margin}
			Comparison of the unexpectedness evaluated by \algoname
			with equation (\ref{eqn:sums}) 
			over the different labels obtained from human evaluation.
		}
	\end{center}
\end{figure}


Some complementary information about the behaviour is provided by the
precision-recall graph of Figure~\ref{fig:precision-recall}: first of all,
\algoname, $\AA$, $\MQ$ and their combinations have larger precision than the
remaining ones at almost all the recall levels; on the other hand
\algoname+$\AA$ is the best method for recall values up to $50\%$, and \algoname
has definitely better precision than $\AA$ until $30\%$ of recall.

In fact, $\MQ$, $\AA$ and \algoname seem to be complementary to one another; in
some sense, this is not surprising given that they stem from completely
different sources of information: one is based on the textual content, another
on the pure link graph and the latter on the category data. 

Some further clue on the behaviour of \algoname is provided by
Figure~\ref{fig:rating-vs-margin}, where the
distribution of \algoname expectedness values is shown for
each of the four labels provided by the human evaluation. The red line is the
median and the upper/lower hinge represent the 75th/25th percentile. 

Finally, let us remark that in order to enhance reproducibility and to foster
further research on this problem, we are sharing all code and data needed to
replicate our findings on \url{http://git.io/vmzjm}.

\section{Conclusions and future work}
\label{sec:conclusions}

In this work we presented a graph model based on the interplay between
categories, able to catch the notion of \emph{expected links} on a graph; we
showed that this model can be employed to find unexpected links in hyperlinked
document corpora, through the determination of a latent category matrix; the
latter is built using a perceptron-like technique. We demonstrated that our
method provides better accuracy than most existing text-based techniques, with
higher efficiency and relying on a much smaller amount of information. Moreover,
we showed that \algoname can process graphs with $10^8$
links or more without effort.
We carried out experiments on the categorized Wikipedia graph -- a widely
employed source of information for knowledge representation. It would be
useful to try our unexpected link mining approach on a different graph, but
we could not find any other openly-accessible dataset of expected/unexpected links.

An interesting question is whether the latent category matrix can be used to
improve link prediction \emph{per se}, i.e.~if it is useful to find links and
not only unexpected ones: this problem requires that one finds a way to bypass
the generalization effect that the matrix produces (for example, by introducing
stochastic behaviour in our model).
Another possible direction would be to try different approach to the
classification problem described in Section~\ref{sec:learningcatmat}, in order
to improve its effectiveness. To this aim, one could recast the problem as a
cost-sensitive classification where false negatives are more costly than false
positives. Other useful techniques include active
learning~\cite{activelearning}: since we need a subset of the non-linked pairs
as counter-examples, active learning would select the more effective ones. An
alternative approach to the same task would be to employ one-class
learning~\cite{oneclasssurvey}. This is left as future work.

\smallskip
\noindent {\bf Acknowledgments}

Authors acknowledge the EU-FET grant NADINE (GA 288956). They also would like to
thank Sebastiano Vigna, both for useful discussions and for providing some
effective code to parse the Wikipedia snapshot. 

{
\bibliography{special-short}
}

\end{document}